\documentclass[12pt]{amsart}
\usepackage{mathrsfs}
\usepackage{amssymb} %Extra symbols 
\usepackage{url}
\usepackage{hyperref}
\usepackage{graphicx} %Include postscript graphics
\usepackage{tikz}
\usetikzlibrary{matrix,arrows}

\theoremstyle{definition}

\newcommand{\image}[3]{
\begin{center}
\begin{figure*}[ht]
\includegraphics[width=#2\textwidth]{#1}
\caption{\small{\label{#1}#3}}\end{figure*}
\end{center}
}

\def\({\left(}
\def\){\right)}

\newcommand{\R}{\mathbb{R}}

\newcommand{\de}{\textnormal{d}}

\newcommand{\tn}{\textnormal}
\newcommand{\ds}{\displaystyle}

\newcommand{\ie}{\textit{i.e.} }
\newcommand{\vs}{\textit{vs.} }

\newcommand{\etc}{\textit{etc.}}

\newcommand{\citep}[2]{\cite{#1}, p. #2}

\newcommand{\IM}{\tn{im }}

\newcommand{\mf}[1]{\mathfrak{#1}}
\newcommand{\mc}[1]{\mathcal{#1}}
\newcommand{\ms}[1]{\mathscr{#1}}

\newcommand{\abs}[1]{\left|#1\right|}
\newcommand{\rank}{\textnormal{rank }}

\newcommand{\ric}{\tn{Ric}}

\newcommand{\dsfrac}[2]{\ds{\frac{#1}{#2}}}

\newcommand{\idxannih}[2]{#1{}^{#2}{}}
\newcommand{\idxcoannih}[2]{#1{}_{#2}{}}

\newcommand{\radix}[1]{\idxcoannih{#1}{\circ}}
\newcommand{\annih}[1]{\idxannih{#1}{\bullet}}
\newcommand{\coannih}[1]{\idxcoannih{#1}{\bullet}}
\newcommand{\coradix}[1]{\idxannih{#1}{\circ}}

\newcommand{\annihg}{\coannih{g}}
\newcommand{\coannihg}{\annih{g}}
\newcommand{\metric}[1]{\langle#1\rangle}
\newcommand{\annihprod}[1]{\coannih{\langle\!\langle#1\rangle\!\rangle}}

\newcommand{\cocontr}{{{}_\bullet}}

\newcommand{\vectmodule}{\mf X}
\newcommand{\fivect}[1]{\vectmodule(#1)}

\newcommand{\fiscal}[1]{\ms F(#1)}

\newcommand{\annihforms}[1]{\annih{\mc A}(#1)}

\newcommand{\kosz}{\mc K}
\newcommand{\der}{\nabla}
\newcommand{\dera}[1]{\der_{#1}}
\newcommand{\derb}[2]{\dera{#1}{#2}}
\newcommand{\derc}[3]{({\derb{#1}{#2}})(#3)}
\newcommand{\lder}{\der^{\flat}}
\newcommand{\ldera}[1]{\lder_{#1}}
\newcommand{\lderb}[2]{\ldera{#1}{#2}}
\newcommand{\lderc}[3]{(\lderb{#1}{#2})(#3)}

\def\hyph{-\penalty0\hskip0pt\relax}
\newcommand{\semiriem}{semi{\hyph}Riemannian}

\newcommand{\semireg}{semi{\hyph}regular}
\newcommand{\ssemireg}{Semi{\hyph}regular}
\newcommand{\quasireg}{quasi{\hyph}regular}

\newcommand{\nondeg}{non{\hyph}degenerate}

\newcommand{\flrw}{Friedmann-Lema\^itre-Robertson-Walker}
\newcommand{\FLRW}{FLRW}
\newcommand{\schw}{Schwarzschild}
\newcommand{\rn}{Reissner-Nordstr\"om}
\newcommand{\kn}{Kerr-Newman}
\newcommand{\hor}{Ho{\v{r}}ava}
\newcommand{\HL}{\hor-Lifschitz}

\begin{document} 

%--------------------------------------------------------
% Title
\title{An Exploration of the Singularities in General Relativity}

\author{Cristi \ Stoica}
\date{May 15, 2012. To the memory of my father, whose birthday is today.}

\begin{abstract}
In General Relativity, spacetime singularities raise a number of problems, both mathematical and physical.

One can identify a class of singularities -- with smooth but degenerate metric -- which, under a set of conditions, allow us to define proper geometric invariants, and to write field equations, including equations which are equivalent to Einstein's at non-singular points, but remain well-defined and smooth at singularities. This class of singularities is large enough to contain isotropic singularities, warped-product singularities, including the {\flrw} singularities, \etc Also a Big-Bang singularity of this type automatically satisfies Penrose's Weyl curvature hypothesis.

The {\schw}, {\rn}, and {\kn} singularities apparently are not of this benign type, but we can pass to coordinates in which they become benign. The charged black hole solutions {\rn} and {\kn} can be used to model classical charged particles in General Relativity. Their electromagnetic potential and electromagnetic field are analytic in the new coordinates -- they have finite values at $r=0$.

There are hints from Quantum Field Theory and Quantum Gravity that a dimensional reduction is required at small scale. A possible explanation is provided by benign singularities, because some of their properties correspond to a reduction of dimensionality.
\end{abstract}

%--------------------------------------------------------
% Title and contents

\begin{center}
\LARGE
\href{http://www.jinr.ru/news_article.asp?n_id=1246}{Seminar held at JINR, Dubna, May 15, 2012.\footnote{The author expresses his gratitude to P. Fiziev and D.V. Shirkov for the invitation, for the perfect organization of this seminar and the staying, and for the interesting discussions and suggestions.}}
\end{center}
\vspace{0.5in}

\maketitle

\setcounter{tocdepth}{1}
\tableofcontents

%----------- section --------------------------------------------------%
\pagebreak
\section{Introduction}

%----------- subsection --------------------------------------------------%
\subsection{Problems of General Relativity}
There are two big problems in General Relativity:
\begin{enumerate}
	 \item 
	It predicts the occurrence of \textit{singularities} \cite{Pen65,Haw66i,Haw66ii,Haw67iii,HP70}.
	\item 
	The attempts to \textit{quantize gravity} seem to fail, because it is perturbatively nonrenormalizable \cite{HV74qg,GS86uvgr}.
\end{enumerate}

Are these problems signs that we should \textit{give up General Relativity} in favor of more radical approaches (superstrings, loop quantum gravity \etc)?

There is another possibility: the limits may be in fact not of GR, but of our tools. Understanding how much we can push the boundaries of GR would be helpful, even in the eventuality that a better theory will replace GR.

%----------- subsection --------------------------------------------------%
\subsection{Two types of singularities}

There are two types of singularities:

\begin{enumerate}
	\item \textit{Malign singularities}: some of the components of the metric are divergent: $g_{ab}\to\infty$. 
  \item \textit{Benign singularities}: $g_{ab}$ are smooth and finite, but $\det g\to 0$.
\end{enumerate}

Benign singularities turn out to be, in many cases, manageable \cite{Sto11a,Sto11b,Sto12b}. Important malign singularities turn out to be in fact benign, but they appear to be malign because they are represented in singular coordinates \cite{Sto11e,Sto11f,Sto11g}.

%----------- subsection --------------------------------------------------%
\subsection{What is wrong with singularities}

If some of metric's components are divergent (as in the case of malign singularities), everything seems to be wrong. If the metric is smooth, but its determinant $\det g\to 0$, the usual Riemannian invariants blow up. For example, the covariant derivative can't be defined, because the inverse of the metric, $\textcolor{red}{g^{ab}}$, becomes singular ($g^{ab}\to\infty$ when $\det g\to 0$). This makes the Christoffel's symbols of the second kind singular:
\begin{equation}
		\Gamma^{\textcolor{red}{c}}{}_{ab} = \ds{\frac 1 2} \textcolor{red}{g^{cs}}(
		\partial_a g_{bs} + \partial_b g_{sa} - \partial_s g_{ab})
\end{equation}
The Riemann curvature is singular too:
\begin{equation}
		R^{\textcolor{red}{d}}{}_{abc} = \Gamma^{\textcolor{red}{d}}{}_{ac,b} - \Gamma^{\textcolor{red}{d}}{}_{ab,c} + \Gamma^{\textcolor{red}{d}}{}_{bs}\Gamma^{\textcolor{red}{s}}{}_{ac} - \Gamma^{\textcolor{red}{d}}{}_{cs}\Gamma^{\textcolor{red}{s}}{}_{ab}
\end{equation}
In addition, the Einstein tensor becomes singular too:
\begin{equation}
		G_{ab} = R_{ab} - \dsfrac{1}{2}R g_{ab}
\end{equation}
and the Ricci and scalar curvatures too:
\begin{equation}
		R_{ab} = R^{\textcolor{red}{s}}{}_{asb}
\end{equation}
\begin{equation}
		R = \textcolor{red}{g^{pq}}R_{pq}.
\end{equation}

%----------- subsection --------------------------------------------------%
\subsection{What are the non-singular invariants?}

Some quantities which are part of the equations are indeed singular, but this is not a problem if we use instead other quantities, equivalent to them when the metric is {\nondeg} \cite{Sto11a,Sto12b}. In the table \ref{table_invariants} one can see that, if the metric is {\nondeg}, the Christoffel symbols of the first kind are equivalent to those of the second kind, the Riemann curvature $R^{a}_{bcd}$ is equivalent to $R_{abcd}$, the Ricci and scalar curvatures are equivalent to their densitized versions and to their Kulkarni-Nomizu products (see equation \ref{eq_kulkarni_nomizu}) with the metric.

\begin{table}[htdp]
	\begin{center}
		 \begin{tabular}{ l | l | l}
			 \hline
			 Singular & Non-Singular & When g is... \\ \hline \hline
			 $\Gamma^c{}_{ab}$ (2-nd) & $\Gamma_{abc}$ (1-st) & smooth \\ \hline
			 $R^d{}_{abc}$ & $R_{abcd}$ & {\semireg} \\ \hline
			 $R_{ab}$ & $R_{ab}\sqrt{\abs{\det g}}^W,\,W\leq 2$ & {\semireg} \\ \hline
			 $R$ & $R\sqrt{\abs{\det g}}^W,\,W\leq 2$ & {\semireg} \\ \hline
			 $\ric$ & $\ric \circ g$ & {\quasireg} \\ \hline
			 $R$ & $R g \circ g$ & {\quasireg} \\ \hline
			 \hline
		 \end{tabular}
	\end{center}
	\caption{Singular invariants and their non-singular equivalents.}
	\label{table_invariants}
\end{table}

%----------- section --------------------------------------------------%
\section{The mathematics of singularities}

%----------- section --------------------------------------------------%
\subsection{Degenerate inner product - algebraic properties}

%----------- subsection --------------------------------------------------%
\image{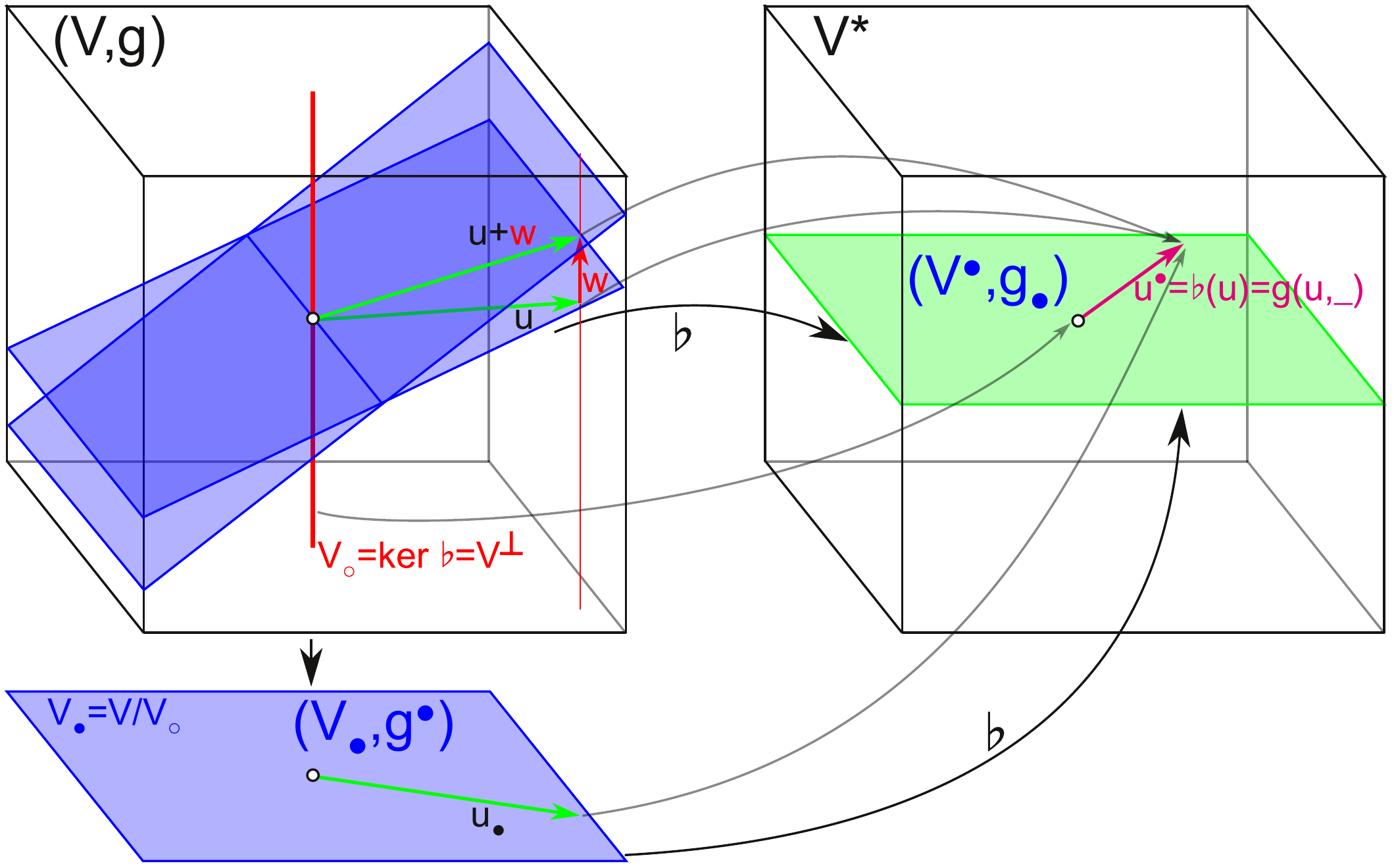}{0.75}{Various vector spaces and inner products associated with a degenerate inner product space $(V,g)$.}

In figure \ref{degenerate-metric}, $(V,g)$ is an inner product vector space. The morphism $\flat:V\to V^*$ is defined by $u\mapsto \annih{u}:=\flat(u)=u^\flat=g(u,\_)$. The radical $\radix{V}:=\ker\flat=V^\perp$ is the set of isotropic vectors in $V$. $\annih{V}:=\IM{\flat}\leq V^*$ is the image of $\flat$. The inner product $g$ induces on $\annih{V}$ an inner product defined by $\annihg(u_1^\flat,u_1^\flat):=g(u_1,u_2)$, which is the inverse of $g$ iff $\det g\neq 0$. The quotient $\coannih{V}:=V/\radix{V}$ consists in the equivalence classes of the form $u+\radix{V}$. On $\coannih{V}$, $g$ induces an inner product $\coannihg(u_1+\radix{V},u_2+\radix{V}):=g(u_1,u_2)$.

%----------- subsection --------------------------------------------------%
\subsection{Relations between the various spaces}

The relations between the radical, the radical annihilator and the factor spaces can be collected in the diagram \cite{Sto11i}:

\begin{center}
\begin{tikzpicture}
\matrix (m) [matrix of math nodes, row sep=4em,
column sep=4em, text height=1.5ex, text depth=0.25ex]
{ 0  & \radix{V} & (V,g) & (\coannih{V},\coannihg) & 0 \\
0	& \coradix{V} & V^*	& (\annih{V},\annihg) & 0	\\ };
\path[right hook->]
(m-1-1) edge (m-1-2)
(m-1-2) edge node[auto] {$\radix{i}$}(m-1-3);
\path[->>]
(m-1-3) edge node[auto] {$\coannih{\pi}$}(m-1-4)
(m-1-4) edge(m-1-5);
\path[->>]
(m-2-2) edge (m-2-1)
(m-2-3) edge node[auto,swap] {$\coradix{\pi}$}(m-2-2)
(m-1-3) edge node[auto,sloped] {$\flat_V$}(m-2-4);
\path[left hook->]
(m-2-4) edge node[auto,swap] {$\annih{i}$}(m-2-3)
(m-2-5) edge(m-2-4);
\path[->]
(m-1-4) edge[bend right=15] node[auto,left] {$\flat$} (m-2-4)
(m-2-4) edge[bend right=15] node[auto,right]{$\sharp$} (m-1-4);
\end{tikzpicture}
\end{center}
where $\coannih{V}=\annih{V}^*=\dsfrac V{\radix V}$ and $\coradix{V}=\radix{V}^*=\frac {V^*}{\annih V}$.

%----------- section --------------------------------------------------%
\subsection{Covariant derivative}

%----------- subsection --------------------------------------------------%
\subsubsection{The Koszul form}

\textit{The Koszul form} is defined as $\kosz:\fivect M^3\to\R$,
\begin{equation}
\label{eq_Koszul_form}
\begin{array}{llll}
	\kosz(X,Y,Z) &:=&\ds{\frac 1 2} \{ X \metric{Y,Z} + Y \metric{Z,X} - Z \metric{X,Y} \\
	&&\ - \metric{X,[Y,Z]} + \metric{Y, [Z,X]} + \metric{Z, [X,Y]}\}.
\end{array}
\end{equation}

In local coordinates it is the Christoffel's symbols of the first kind:
\begin{equation}
\label{eq_Koszul_form_coord}
	\kosz_{abc}=\kosz(\partial_a,\partial_b,\partial_c)=\ds{\frac 1 2} (
	\partial_a g_{bc} + \partial_b g_{ca} - \partial_c g_{ab}) = \Gamma_{abc},
\end{equation}

For {\nondeg} metrics, the Levi-Civita connection is obtained uniquely:
\begin{equation}
\label{eq_koszul_formula_inv}
	\textcolor{red}{\derb X Y} = \kosz(X,Y,\_)^{\textcolor{red}{\sharp}}.
\end{equation}

For degenerate metrics, we will have to avoid the usage of the Levi-Civita connection, and limit ourselfs to using the Koszul form as much as possible.

%----------- subsection --------------------------------------------------%
\subsubsection{The covariant derivatives}
The \textit{lower covariant derivative} of a vector field $Y$ in the direction of a vector field $X$ \cite{Sto11a}:
\begin{equation}
\label{eq_l_cov_der_vect}
\lderc XYZ := \kosz(X,Y,Z)
\end{equation}

If the Koszul form satisfies the condition that $\kosz(X,Y,W)=0$ whenever the vector field $W$ satisfies $W\in\Gamma(\radix{T}M)$, the singular {\semiriem} manifold $(M,g)$ is named \textit{radical stationary}.

The covariant derivative of differential forms can be defined :
\begin{equation*}
	\left(\der_X\omega\right)(Y) := X\left(\omega(Y)\right) - \annihprod{\lderb X Y,\omega},
\end{equation*}
if $\omega\in \annihforms{M}:=\Gamma(\annih{T}M)$. More general,

\begin{equation*}
	\der_X(\omega_1\otimes\ldots\otimes\omega_s) := \der_X(\omega_1)\otimes\ldots\otimes\omega_s +\ldots + \omega_1\otimes\ldots\otimes\der_X(\omega_s)
\end{equation*}

If $T\in\Gamma(\otimes^k_M\annih{T}M)$, then the covariant derivative is
\begin{equation*}
\begin{array}{lll}
	\left(\nabla_X T\right)(Y_1,\ldots,Y_k) &=& X\left(T(Y_1,\ldots,Y_k)\right) \\
	&&- \sum_{i=1}^k\kosz(X,Y_i,\cocontr)T(Y_1,,\ldots,\cocontr,\ldots,Y_k)
\end{array}
\end{equation*}

%----------- subsection --------------------------------------------------%
\subsection{Riemann curvature tensor. {\ssemireg} manifolds.}

The \textit{Riemann curvature tensor} can be defined, for a radical stationary manifold $(M,g)$, by:
\begin{equation}
\label{eq_riemann_curvature_explicit}
	R(X,Y,Z,T) = \derc X {{\ldera Y}Z}T - \derc Y {{\ldera X}Z}T - \lderc {[X,Y]}ZT
\end{equation}

In local coordinates, it takes the form:
\begin{equation}
	R_{abcd}= \partial_a \kosz_{bcd} - \partial_b \kosz_{acd} + (\kosz_{ac\cocontr}\kosz_{bd\cocontr} - \kosz_{bc\cocontr}\kosz_{ad\cocontr})
\end{equation}

The Riemann curvature is a tensor field. It has the same symmetry properties as for $\det g\neq 0$. It is radical-annihilator in each of its slots.

A singular {\semiriem} manifold is called {\semireg} \cite{Sto11a} if:
\begin{equation}
	\dera X {\ldera Y}Z \in \annihforms M.
\end{equation}

Equivalently,
\begin{equation}
	\kosz(X,Y,\cocontr)\kosz(Z,T,\cocontr) \in \fiscal M.
\end{equation}

The Riemann curvature is smooth for {\semireg} metrics.

%~~~~~~~~~~~~~~~~~~~~~~~~~~~~~~~~~~~~~~~~~~~~~~~~~~~~~~~~~~~~~~~~~~~~~~~%
\subsection{Examples of {\semireg} {\semiriem} manifolds}
\label{s_semi_reg_semi_riem_man_example}

The following are examples of \cite{Sto11a,Sto11b}.

%~~~~~~~~~~~~~~~~~~~~~~~~~~~~~~~~~~~~~~~~~~~~~~~~~~~~~~~~~~~~~~~~~~~~~~~%
\subsubsection{Isotropic singularities}

Isotropic singularities have the form
$$g=\Omega^2\tilde g,$$
where $\tilde g$ is a {\nondeg} bilinear form on $M$.

%~~~~~~~~~~~~~~~~~~~~~~~~~~~~~~~~~~~~~~~~~~~~~~~~~~~~~~~~~~~~~~~~~~~~~~~%
\subsubsection{Degenerate warped products}

Degenerate warped products $B\times_f F$ are defined similarly to the usual warped products:
\begin{equation}
\label{eq_wp_metric}
	\de s^2=\de s_B^2 + f^2(p)\de s_F^2.
\end{equation}
The difference from the {\nondeg} case is that for the degenerate warped products, $f$ allowed to vanish. We can take the manifolds $B$ and $F$ to be radical stationary, and the warped product will also be radical stationary, if $\de f\in  \annihforms$. If $B$ and $F$ are {\semireg}, and $\de f\in \annihforms$, but also $\nabla_X\de f\in  \annihforms$ for any vector field $X$, then $B\times_f F$ is {\semireg}.

%~~~~~~~~~~~~~~~~~~~~~~~~~~~~~~~~~~~~~~~~~~~~~~~~~~~~~~~~~~~~~~~~~~~~~~~%
\subsubsection{{\FLRW} spacetimes}

{\FLRW} spacetimes are degenerate warped products:
\begin{equation}
	\de s^2 = -\de t^2 + a^2(t)\de\Sigma^2
\end{equation}
\begin{equation}
\label{eq_flrw_sigma_metric}
\de\Sigma^2 = \dsfrac{\de r^2}{1-k r^2} + r^2\(\de\theta^2 + \sin^2\theta\de\phi^2\),
\end{equation}
where $k=1$ for $S^3$, $k=0$ for $\R^3$, and $k=-1$ for $H^3$.

%~~~~~~~~~~~~~~~~~~~~~~~~~~~~~~~~~~~~~~~~~~~~~~~~~~~~~~~~~~~~~~~~~~~~~~~%
\section{Einstein's equation on {\semireg} spacetimes}
\label{s_einstein_tensor_densitized}

%----------- subsection --------------------------------------------------%
\subsection{Einstein's equation on {\semireg} spacetimes}

On $4$D {\semireg} spacetimes Einstein tensor density $G\det g$ is smooth \cite{Sto11a}:

\begin{equation}
\label{eq_einstein:densitized}
	G\det g + \Lambda g\det g = \kappa T\det g,
\end{equation}

or, in coordinates or local frames,
\begin{equation}
\label{eq_einstein_idx:densitized}
	G_{ab}\det g + \Lambda g_{ab}\det g = \kappa T_{ab}\det g,
\end{equation}

It is not allowed to divide by $\det g$, when $\det g=0$.

%~~~~~~~~~~~~~~~~~~~~~~~~~~~~~~~~~~~~~~~~~~~~~~~~~~~~~~~~~~~~~~~~~~~~~~~%
\section{{\flrw} spacetime}

%----------- subsection --------------------------------------------------%
\subsection{The {\flrw} spacetime}

If $S$ is a connected three-dimensional Riemannian manifold of constant curvature $k\in\{-1,0,1\}$ (\ie $H^3$,$\R^3$ or $S^3$) and $a\in (A,B)$, $-\infty\leq A < B \leq \infty$, $a\geq 0$, then the warped product $I \times_a S$ is called a \textit{Friedmann-Lema\^itre-Robertson-Walker} spacetime.

\begin{equation}
	\de s^2 = -\de t^2 + a^2(t)\de\Sigma^2
\end{equation}
\begin{equation}
\de\Sigma^2 = \dsfrac{\de r^2}{1-k r^2} + r^2\(\de\theta^2 + \sin^2\theta\de\phi^2\),
\end{equation}
where $k=1$ for $S^3$, $k=0$ for $\R^3$, and $k=-1$ for $H^3$.

In general the warping function is taken $a\in\fiscal{I}$ is $a>0$. Here we allow it to be $a\geq 0$, including possible singularities.

The resulting singularities are {\semireg}.

%----------- subsection --------------------------------------------------%
\subsection{Distance separation \vs topological separation}
\image{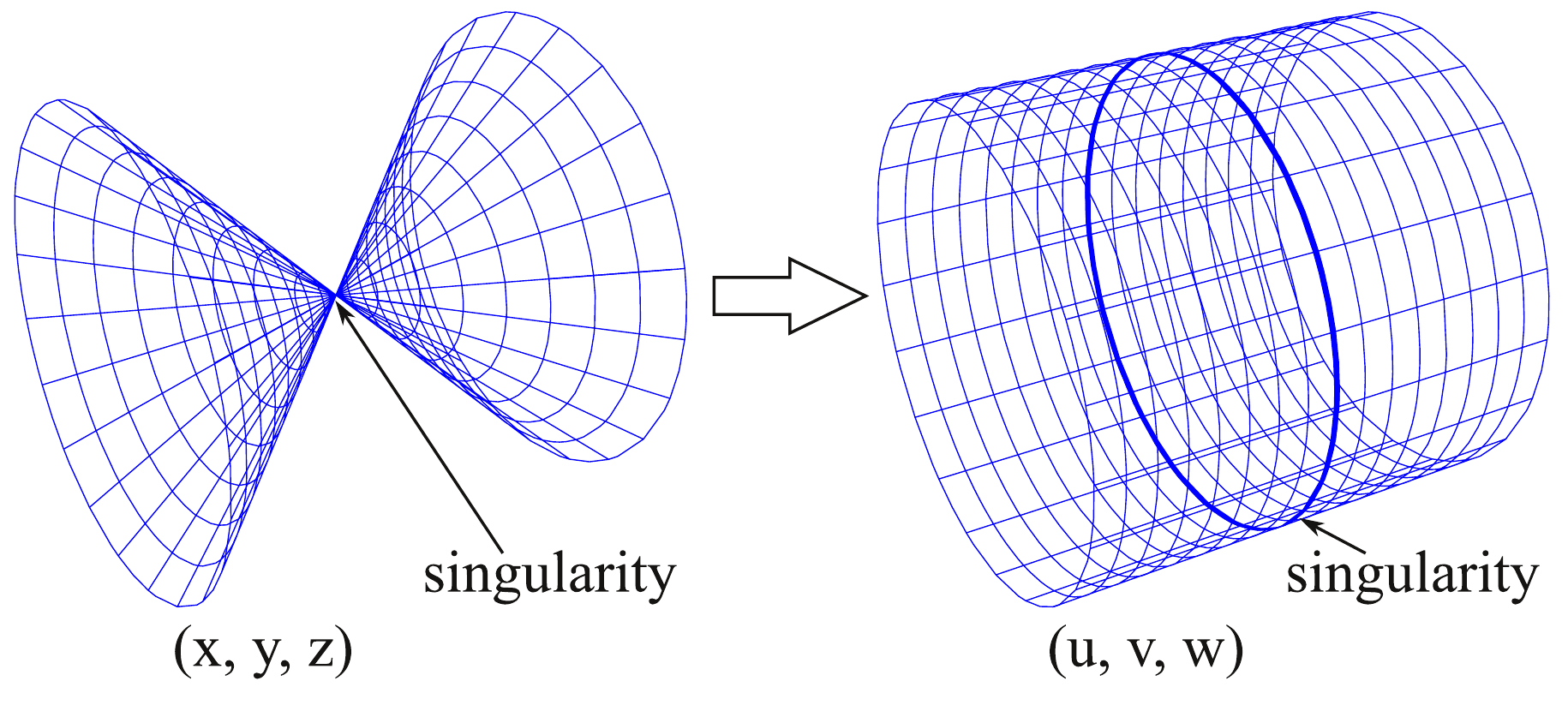}{0.8}{
The old method of resolution of singularities shows how we can ``untie'' the singularity of a cone and obtain a cylinder. Similarly, it is not necessary to assume that, at the Big-Bang singularity, the entire space was a point, but only that the space metric was degenerate.}

%----------- subsection --------------------------------------------------%
\subsection{Friedman equations}
The stress-energy tensor for a fluid in thermodynamic equilibrium, of $\rho$ mass density and $p$ pressure density, is
\begin{equation}
\label{eq_friedmann_stress_energy}
T^{ab} = \(\rho+p\)u^a u^b + p g^{ab},
\end{equation}
where $u^a$ is the timelike vector field $\partial_t$, normalized.

The following equations follow from the above stress-energy tensor, in the case of a homogeneous and isotopic universe.

The \textit{Friedmann equation} is:
\begin{equation}
\label{eq_friedmann_density}
\rho = \dsfrac{3}{\kappa}\dsfrac{\dot{a}^2 + k}{a^2}.
\end{equation}

The \textit{acceleration equation} is:
\begin{equation}
\label{eq_acceleration}
\rho + 3p = -\dsfrac{6}{\kappa}\dsfrac{\ddot{a}}{a}.
\end{equation}

The \textit{fluid equation}, expressing the conservation of mass-energy, is:
\begin{equation}
\label{eq_fluid}
\dot{\rho} = -3 \dsfrac{\dot{a}}{a}\(\rho + p\).
\end{equation}

They are singular for $a=0$.

%----------- subsection --------------------------------------------------%
\subsection{Friedman equations, densitized}

\cite{Sto11h}

The actual densities contain in fact $\sqrt{-g}(=a^3 \sqrt{g_{\Sigma}})$:

\begin{equation}
\label{eq_substitution_densities}
\begin{array}{l}
\bigg\{
\begin{array}{ll}
	\widetilde\rho = \rho \sqrt{-g} = \rho a^3 \sqrt{g_{\Sigma}} \\
	\widetilde p = p \sqrt{-g} = p a^3 \sqrt{g_{\Sigma}} \\
\end{array}
\\
\end{array}
\end{equation}

The Friedmann equation \eqref{eq_friedmann_density} becomes
\begin{equation}
\label{eq_friedmann_density_tilde}
\widetilde\rho = \dsfrac{3}{\kappa}a\(\dot a^2 + k\) \sqrt{g_{\Sigma}},
\end{equation}

The acceleration equation \eqref{eq_acceleration} becomes
\begin{equation}
\label{eq_acceleration_tilde}
\widetilde\rho + 3\widetilde p = -\dsfrac{6}{\kappa}a^2\ddot{a} \sqrt{g_{\Sigma}},
\end{equation}

Hence, $\widetilde \rho$ and $\widetilde p$ are smooth{}, as it is the densitized stress-energy tensor
\begin{equation}
T_{ab}\sqrt{-g} = \(\widetilde\rho+\widetilde p\)u_a u_b + \widetilde p g_{ab}.
\end{equation}

Figure \ref{flrw-def-pos} represents a {\FLRW} Big-Bang singularity for $\dot a(0)>0$. Figure \ref{flrw-def-null} represents the case when $\dot a(0)=0$,  $\ddot a(0)>0$.

\pagebreak
\image{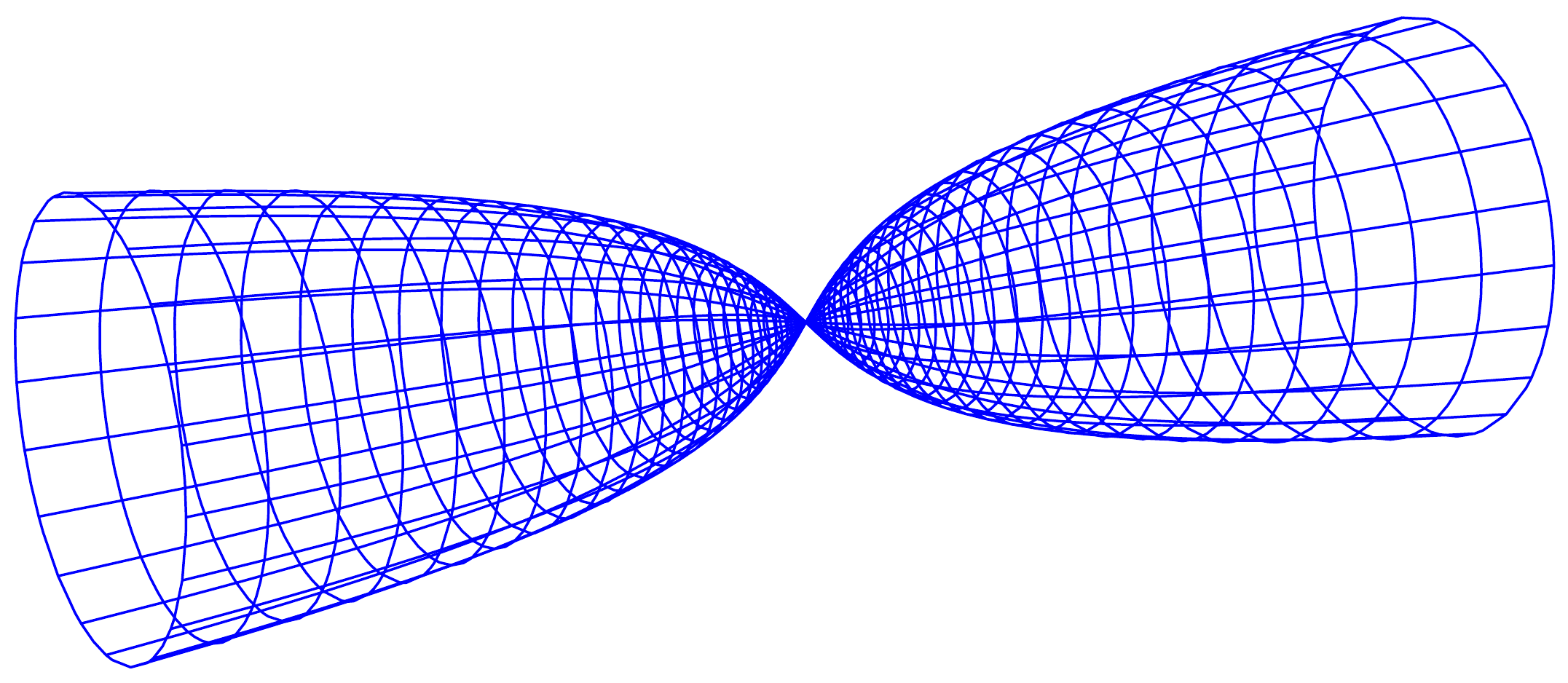}{0.8}{Big Bang singularity, corresponding to $a(0)=0$,  $\dot a(0)>0$.}

\image{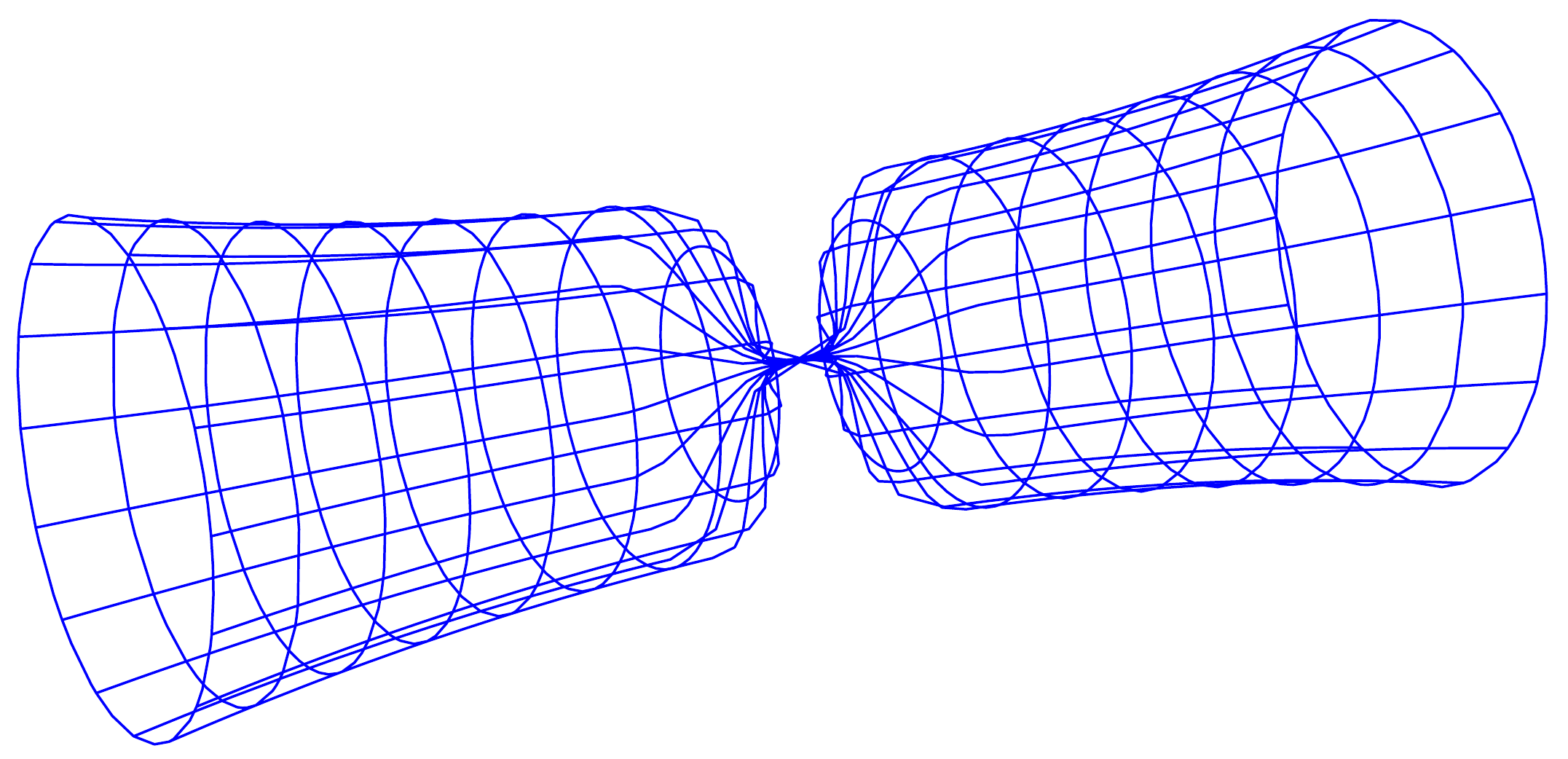}{0.8}{Big Bounce, corresponding to $a(0)=0$,  $\dot a(0)=0$,  $\ddot a(0)>0$.}

\pagebreak
%~~~~~~~~~~~~~~~~~~~~~~~~~~~~~~~~~~~~~~~~~~~~~~~~~~~~~~~~~~~~~~~~~~~~~~~%
\section{Black hole singularities}

%~~~~~~~~~~~~~~~~~~~~~~~~~~~~~~~~~~~~~~~~~~~~~~~~~~~~~~~~~~~~~~~~~~~~~~~%
\subsection{{\schw} black holes}

%----------- subsection --------------------------------------------------%
\subsection{{\schw} singularity is {\semireg}}

The {\schw} metric is given by:
\begin{equation}
\label{eq_schw_schw}
\de s^2 = -\(1-\dsfrac{2m}{r}\)\de t^2 + \(1-\dsfrac{2m}{r}\)^{-1}\de r^2 + r^2\de\sigma^2,
\end{equation}
where
\begin{equation}
\label{eq_sphere}
\de\sigma^2 = \de\theta^2 + \sin^2\theta \de \phi^2
\end{equation}

Let's change the coordinates to
\begin{equation}
\label{eq_coordinate_semireg}
\begin{array}{l}
\bigg\{
\begin{array}{ll}
r &= \tau^2 \\
t &= \xi\tau^4 \\
\end{array}
\\
\end{array}
\end{equation}

The four-metric becomes:
\begin{equation}
\label{eq_schw_analytic_tau_xi}
\de s^2 = -\dsfrac{4\tau^4}{2m-\tau^2}\de \tau^2 + (2m-\tau^2)\tau^{4}\(4\xi\de\tau + \tau\de\xi\)^2 + \tau^4\de\sigma^2
\end{equation}
which is analytic and {\semireg}  at $r=0$ \cite{Sto11e}.

This solution can be foliated in space+time, and can therefore be used to represent evaporating black holes of {\schw} type. Since the solution can be analytically extended beyond the singularity, the information is not lost there (fig. \ref{evaporating-bh-s}).

\image{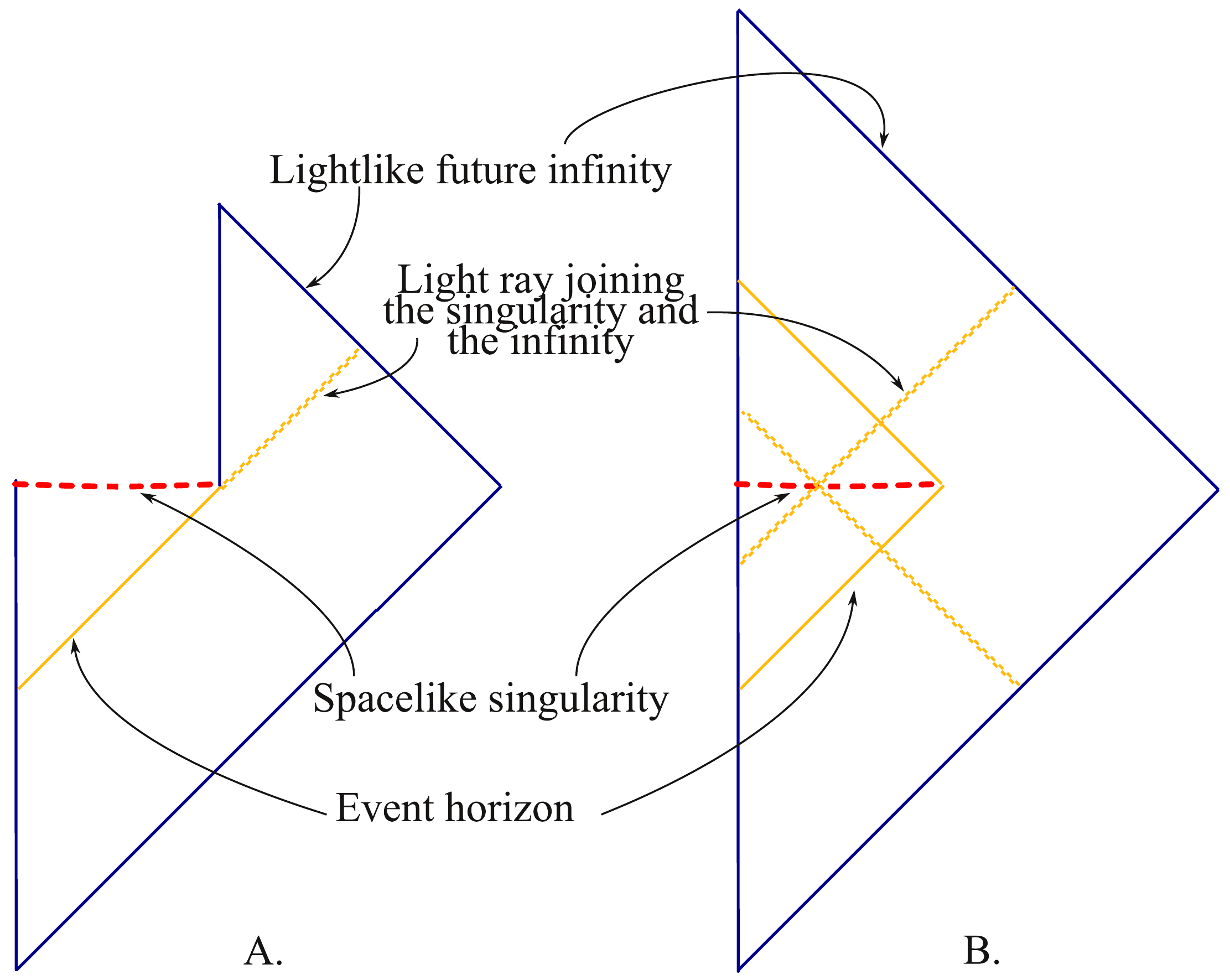}{0.6}{\textbf{A.} Standard evaporating black hole, whose singularity destroys the information.
\textbf{B.} Evaporating black hole extended through the singularity preserves information.}

%~~~~~~~~~~~~~~~~~~~~~~~~~~~~~~~~~~~~~~~~~~~~~~~~~~~~~~~~~~~~~~~~~~~~~~~%
\subsection{{\rn} black holes}

The {\rn} metric is given by:
\begin{equation}
\label{eq_rn_metric}
\de s^2 = -\left(1-\dsfrac{2m}{r} + \dsfrac{q^2}{r^2}\right)\de t^2 + \left(1-\dsfrac{2m}{r} + \dsfrac{q^2}{r^2}\right)^{-1}\de r^2 + r^2\de\sigma^2,
\end{equation}

We choose the coordinates $\rho$ and $\tau$ \cite{Sto11f}, so that
\begin{equation}
	\begin{array}{l}
	\bigg\{
	\begin{array}{ll}
	t &= \tau\rho^T \\
	r &= \rho^S \\
	\end{array}
	\\
	\end{array}
\end{equation}

The metric has, in the new coordinates, the following form
\begin{equation}
\label{eq_rn_ext_ext}
\de s^2 = - \Delta\rho^{2T-2S-2}\left(\rho\de\tau + T\tau\de\rho\right)^2 + \dsfrac{S^2}{\Delta}\rho^{4S-2}\de\rho^2 + \rho^{2S}\de\sigma^2,
\end{equation}
\begin{equation}
	\tn{where \,\,}\Delta := \rho^{2S} - 2m \rho^{S} + q^2.
\end{equation}

To remove the infinity of the metric at $r=0$, take
\begin{equation}
	\begin{array}{l}
	\bigg\{
	\begin{array}{ll}
	S \geq 1 \\
	T \geq S + 1	
	\end{array}
	\\
	\end{array}
\end{equation}

which also ensure that the metric is analytic at $r=0$.

The electromagnetic potential in the coordinates $(t,r,\phi,\theta)$ is singular at $r=0$ \cite{Sto11f}:
\begin{equation}
A = -\dsfrac q r \de t,
\end{equation}

In the new coordinates $(\tau,\rho,\phi,\theta)$, the electromagnetic potential is
\begin{equation}
A = -q\rho^{T-S-1}\left(\rho\de\tau + T\tau\de\rho\right),
\end{equation}

the electromagnetic field is
\begin{equation}
F = q(2T-S)\rho^{T-S-1}\de\tau \wedge\de\rho,
\end{equation}

and they are analytic everywhere, including at the singularity $\rho=0$.

To have space+time foliation given by the coordinate, must have $T\geq 3S$ \cite{Sto11f}.

\image{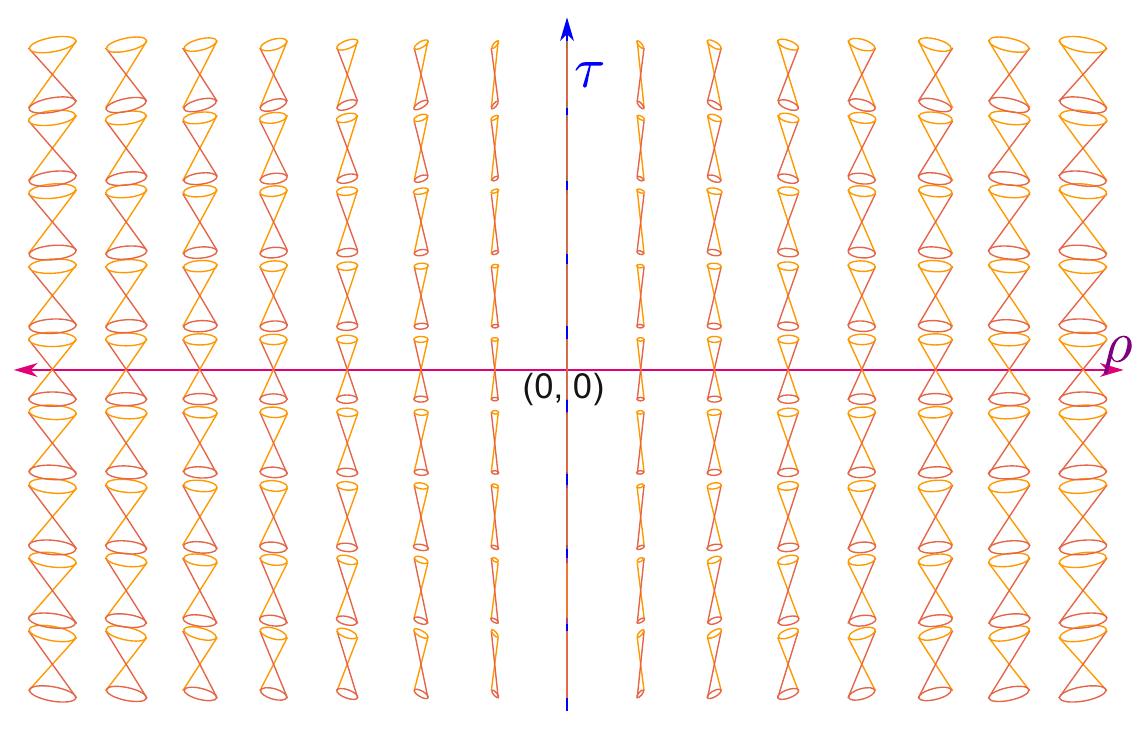}{0.68}{As one approaches the singularity on the axis $\rho=0$, the lightcones become more and more degenerate along that axis (for $T\geq 3S$ and even $S$).}

%----------- section --------------------------------------------------%
\section{Global hyperbolicity and information}

%~~~~~~~~~~~~~~~~~~~~~~~~~~~~~~~~~~~~~~~~~~~~~~~~~~~~~~~~~~~~~~~~~~~~~~~%
\subsection{Foliations with Cauchy hypersurfaces}

If the singularities are benign, the evolution equations can make sense. But to be able to formulate initial value problems, it is needed that spacetime admits space+time foliations with respect to the metric tensor. The spacelike hypersurfaces have to be Cauchy surfaces, which is equivalent to the global hyperbolicity condition. The spacelike hypersurfaces must have the same topology for any moment of time $t$, but the metric may change its rank, and become sometimes degenerate. As we will see, the stationary black hole singularities are compatible with such foliations, hence with global hyperbolicity. Also because the topology seems to be independent on the quantities $m$, $q$, and $a$ which characterize the black hole, being their only ``hair'', it is possible by varying these quantities to construct models of black holes which appear and disappear. This is relevant to evaporating black holes, since now we can see that the information is not necessarily lost.

%----------- subsection --------------------------------------------------%
\subsection{Space-like foliation of the {\rn} solution}

We can foliate the {\rn} solution in Cauchy hypersurfaces, if it is expressed in the new coordinates \cite{Sto11f,Sto11c,Sto12e}. Figure \ref{up-big} illustrates such a situation.

\image{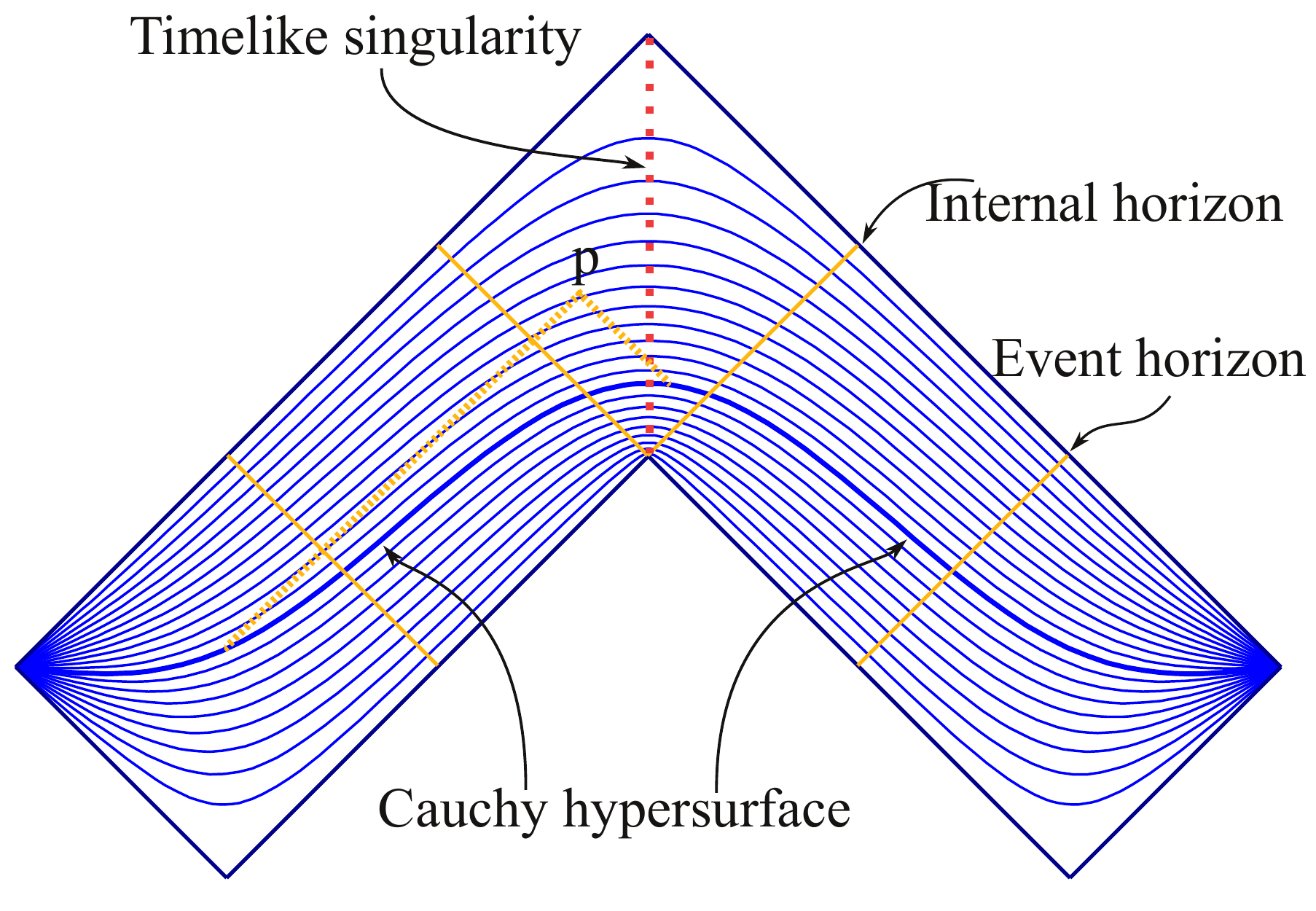}{0.63}{Foliation of the {\rn} solution in Cauchy hypersurfaces. The spacetime regions beyond the Cauchy horizons are ignored, to obtain a global foliation.}

Similar foliations were obtained for the extremal and naked {\rn} black holes, as well as for the {\schw} and {\kn} black holes.

Implications: we can vary $m$, $q$, $a$ and obtain general singularities which preserve information.

%----------- section --------------------------------------------------%
\section{The mathematics of singularities 2}

%----------- subsection --------------------------------------------------%
\subsection{The Ricci decomposition}

The Riemann curvature tensor can be decomposed algebraically as
\begin{equation}
	R_{abcd} = S_{abcd} + E_{abcd} + C_{abcd}
\end{equation}
where 
\begin{equation}
	S_{abcd} = \dsfrac{1}{n(n-1)}R(g\circ g)_{abcd}
\end{equation}
\begin{equation}
	E_{abcd} = \dsfrac{1}{n-2}(S \circ g)_{abcd}
\end{equation}
\begin{equation}
\label{eq_ricci_traceless}
S_{ab} := R_{ab} - \dsfrac{1}{n}Rg_{ab}
\end{equation}
where the Kulkarni-Nomizu product is used:
\begin{equation}
\label{eq_kulkarni_nomizu}
	(h\circ k)_{abcd} := h_{ac}k_{bd} - h_{ad}k_{bc} + h_{bd}k_{ac} - h_{bc}k_{ad}
\end{equation}

If the Riemann curvature tensor on a {\semireg} manifold $(M,g)$ admits such a decomposition which is smooth, $(M,g)$ is said to be \textit{\quasireg}.

%----------- subsection --------------------------------------------------%
\subsection{The expanded Einstein equation}

\cite{Sto12b}

In dimension $n=4$ we introduce the \textit{expanded Einstein equation}
\begin{equation}
\label{eq_einstein_expanded}
	(G\circ g)_{abcd} + \Lambda (g\circ g)_{abcd} = \kappa (T\circ g)_{abcd}
\end{equation}

or, equivalently,
\begin{equation}
\label{eq_einstein_expanded_explicit}
	2 E_{abcd} - 3 S_{abcd} + \Lambda (g\circ g)_{abcd} = \kappa (T\circ g)_{abcd}.
\end{equation}

It is equivalent to Einstein's equation if the metric is {\nondeg}.

%----------- subsection --------------------------------------------------%
\subsection{Examples of {\quasireg} singularities}

\cite{Sto12b}

\begin{itemize}
	\item 
	Isotropic singularities.
	\item 
	Degenerate warped products $B\times_f F$ with $\dim B=1$ and $\dim F=3$.
	\item 
	In particular, {\FLRW} singularities \cite{Sto12a}.
	\item 
	{\schw} singularities.
	\item 
	The question whether the {\rn} and {\kn} singularities are {\semireg}, or {\quasireg}, is still open.
\end{itemize}

%----------- subsection --------------------------------------------------%
\subsection{The Weyl tensor at {\quasireg} singularities}

\cite{Sto12c}

The \textit{Weyl curvature tensor}:
\begin{equation}
\label{eq_weyl_curvature}
	C_{abcd} = R_{abcd} - S_{abcd} - E_{abcd}.
\end{equation}

$C_{abcd} \to 0$ as approaching a {\quasireg} singularity.

Because of this, any {\quasireg} Big Bang satisfies the Weyl curvature hypothesis, emitted by Penrose to explain the low entropy at the Big Bang.

%~~~~~~~~~~~~~~~~~~~~~~~~~~~~~~~~~~~~~~~~~~~~~~~~~~~~~~~~~~~~~~~~~~~~~~~%
\section{Dimensional reduction and QFT}

%----------- subsection --------------------------------------------------%
\subsection{Hints of dimensional reduction in QFT and QG}

Various results obtained in QFT and in Quantum Gravity suggest that at small scales there should take place a dimensional reduction. The explanation of this reduction differs from one approach to another. Examples of such results are given below:
\begin{itemize}
	\item 
	The scattering amplitudes in QCD \cite{Lip88,Lip89,Lip91}.
	\item 
	High energy Regge regime \cite{VV93qcd,Aref94regge}.
	\item 
	Fractal universe \cite{Calc2010FractalQFT,Calc2010FractalUniverse,Calc2011FractalGravity}.
	\item 
	Topological dimensional reduction \cite{shirkov2010coupling,FS2011KG,Fiz2010Riem,FS2012Axial,shirkov2012dreamland}.
	\item 
	Vanishing Dimensions at LHC \cite{ADFLS10}.
	\item 
	Dimensional reduction in Quantum Gravity \cite{Car95,Car09SDR,Car10sssst}.
	\item 
	Asymptotic safety \cite{Wein79AS}.
	\item 
	Causal dynamical triangulations \cite{AJL00,AJL04,AJL05r,AJL05s,AJL09aqg}.
	\item 
	{\HL} gravity \cite{Hor09qglp}.
\end{itemize}

%----------- subsection --------------------------------------------------%
\subsection{Is dimensional reduction due to the benign singularities?}

We will make some connections between these results and the benign singularities \cite{Sto12d}.

	First, at each point where the metric becomes degenerate, a geometric, or \textit{metric reduction} takes place, because the rank of the metric is reduced:
	\begin{equation}
	\dim{\coannih{T_p}M}=\dim{\annih{T_p}M}=\rank g_p.
\end{equation}

	This is in fact a local effect (it goes on an entire neighborhood of that point), in the regions of constant signature of the metric. This follows from the Kupeli theorem \cite{Kup87b}: for constant signature, the space is locally a \textit{product} $M=P\times_0 N$ between a manifold of lower dimension $P$ and another manifold $N$ with metric $0$. This suggests a connection with the \textit{topological dimensional reduction} explored by D.V. Shirkov and P. Fiziev \cite{shirkov2010coupling,FS2011KG,Fiz2010Riem,FS2012Axial,shirkov2012dreamland}.

	If the singularity is {\quasireg}, the Weyl tensor $C_{abcd} \to 0$ as approaching a {\quasireg} singularity. This implies that the \textit{local degrees of freedom vanish}, \ie the gravitational waves for GR and the gravitons for QG \cite{Car95}.

In \cite{Sto11f} we obtained new coordinates, which make the {\rn} metric analytic at the singularity. In these coordinates, the metric is
\begin{equation}
\de s^2 = - \Delta\rho^{2T-2S-2}\left(\rho\de\tau + T\tau\de\rho\right)^2 + \dsfrac{S^2}{\Delta}\rho^{4S-2}\de\rho^2 + \rho^{2S}\de\sigma^2.
\end{equation}
A \textit{charged particle} can be viewed, at least classically, as a {\rn} black hole. The above metric reduces its dimension to dim $=2$.

	To admit space+time foliation in these coordinates, we should take \textit{$T\geq 3S$}. Is this anisotropy connected to \textit{\HL} gravity?

	In the \textit{fractal universe approach} \cite{Calc2010FractalQFT,Calc2010FractalUniverse,Calc2011FractalGravity}, one expresses the measure in
\begin{equation}
S=\int_\mc M\de\varrho(x)\, \mc L
\end{equation}
in terms of some functions $f_{(\mu)}(x)$, some of them vanishing at low scales:
\begin{equation}
\label{eq_stme}
\de \varrho(x) = \prod_{\mu=0}^{D-1} f_{(\mu)}(x)\,\de x^\mu
\end{equation}

In \textit{Singular General Relativity},
\begin{equation}
\label{eq_stme_sgr}
\de \varrho(x) = \sqrt{-\det g}\de x^D,
\end{equation}
If the metric is diagonal in the coordinates $(x^\mu)$, then we can take
\begin{equation}
\label{eq_stme_sgr_weights}
f_{(\mu)}(x) = \sqrt{\abs{g_{\mu\mu}(x)}}.
\end{equation}

This suggests that the results obtained by Calcagni by considering the universe to be fractal follow naturally from the benign metrics.

%----------- subsection --------------------------------------------------%
\subsection{How can dimension vary with scale?}

Dimension vary with distance. But how can it vary with scale? A (still very vague) answer can be the following: metric's average determinant decreases as the number of singularities (particles) in the region increases. We conjecture that this happens and gives the regularization \cite{Sto12d} (see fig. \ref{dim-red-feynman}).

\image{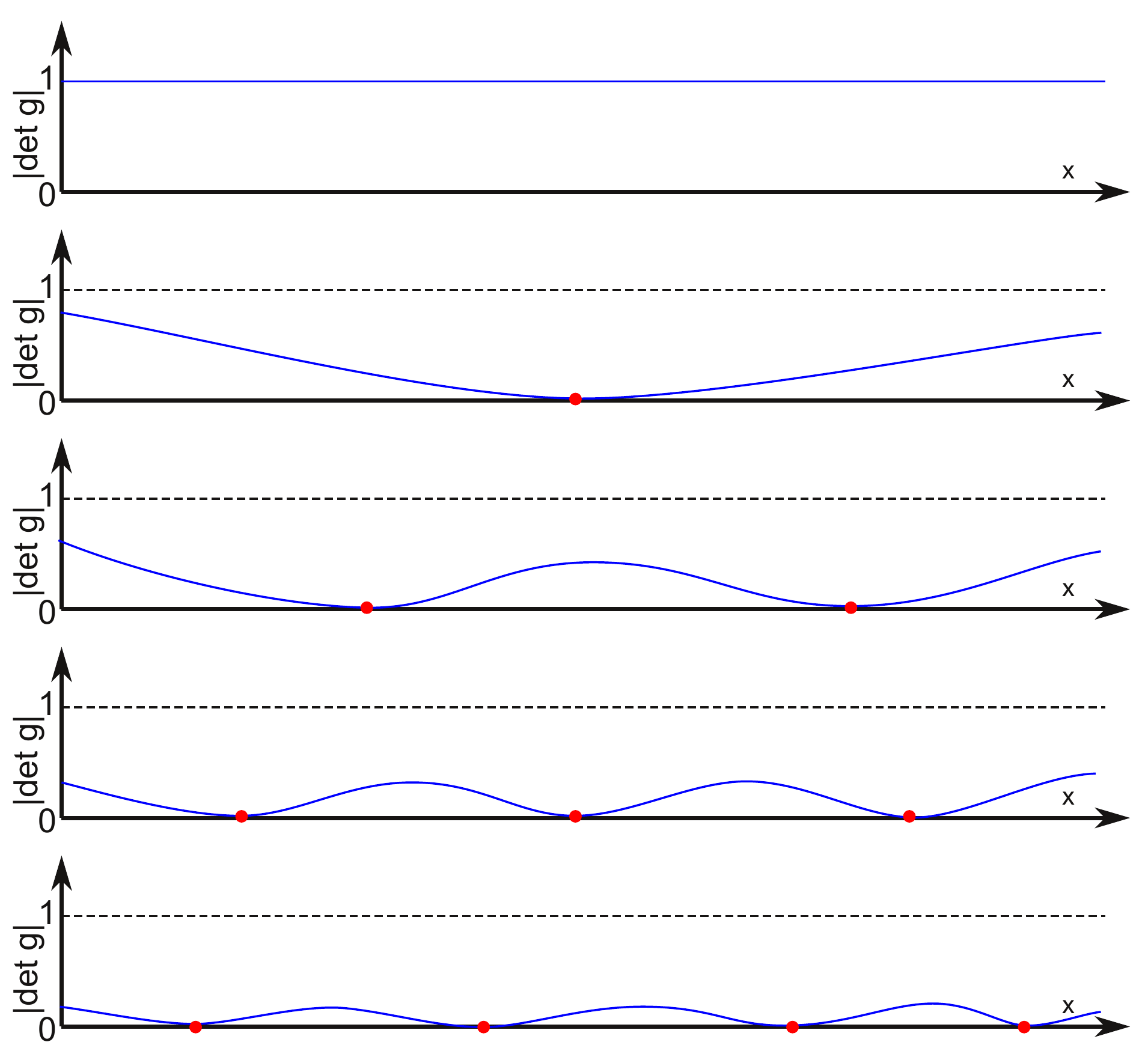}{0.6}{
\textit{A conjecture:} Metric's average determinant decreases as the number of singularities (particles) in the region increases.}

\pagebreak
\bibliographystyle{unsrt}%{unsrt}%{amsalpha}%{amsplain}
%\bibliography{sing-gr_bib}

\end{document}